\documentclass[preprint,prd,showpacs,amssymb,floatfix]{revtex4} 
 
\begin{document} 
\input{epsf} 

\title{Minkowski Vacuum Stress Tensor Fluctuations}

\author{ L.H. Ford}
 \email[Email: ]{ford@cosmos.phy.tufts.edu} 
 \affiliation{Institute of Cosmology  \\
Department of Physics and Astronomy\\ 
         Tufts University, Medford, MA 02155}
\author{Thomas A. Roman}
  \email[Email: ]{roman@ccsu.edu}
  \affiliation{Department of  Mathematical Sciences \\
 Central Connecticut State University \\  
New Britain, CT 06050}  

\begin{abstract}
We study the fluctuations of the stress tensor for a massless scalar field
in two and four-dimensional Minkowski spacetime in the vacuum state. Covariant
expressions for the stress tensor correlation function are obtained as sums
of derivatives of a scalar function. These expressions allow one to express
spacetime averages of the correlation function as finite integrals. We also
study the correlation between measurements of the energy density along a
worldline. We find that these measurements may be either positively correlated
or anticorrelated. The anticorrelated measurements can be interpreted as
telling us that if one measurement yields one sign for the averaged energy
density, a successive measurement with a suitable time delay is likely
to yield a result with the opposite sign.
\end{abstract}

\maketitle

\baselineskip=13pt

\section{Introduction}
\label{sec:intro}

Because physically realizable states in quantum field theory are not
eigenstates of the stress tensor operator, quantum stress tensor fluctuations
are a universal feature of quantum fields. These fluctuations can
have physical effects, including Casimir force 
fluctuations~\cite{Barton,Eberlein,JR,WKF01}, radiation pressure 
fluctuations~\cite{WF01}, and passive fluctuations of the gravitational 
field~\cite{F82,DF88,Kuo,PH97,WF99,CH95,CCV,Moffat,MV99,HS98,BFP00,PH00,FW03,
HRV04,BF,BF2,Winitzki01,Vachaspati03,DV05}. 
Passive fluctuations of gravity are those driven by fluctuations of the matter
field stress tensor, as opposed to the active fluctuations due to the quantum
nature of gravity itself. The quantum stress tensor correlation function
is  singular in the limit of coincident points. However, this does not
prevent us from 
obtaining physically meaningful results for observable quantities,
such as the luminosity fluctuations of a distant source seen through the
fluctuating spacetime~\cite{BF}. These observables are expressed as spacetime
integrals of the correlation function, which can be defined by an integration
by parts procedure. Alternatively, one could use other approaches, such as
dimensional regularization~\cite{FW04}. 

In general, the stress tensor correlation function can be decomposed into
three terms: a ``fully normal ordered'' term which is state dependent, but 
free of singularities, a vacuum term which is singular, but state independent,
and a ``cross term'' which is both singular and state dependent. In many
situations, one is interested in state dependent effects, so the vacuum
term can be ignored. For example, radiation pressure fluctuations in
a coherent state arise solely from the cross term~\cite{WF01}. However,
this does not mean that the vacuum term is devoid of any physical content.

The main purpose of this paper is the search for such content. Here we will
be concerned with a free, massless scalar field in Minkowski spacetime,
and  its stress tensor correlation function in the Minkowski vacuum state.
In a previous paper~\cite{FR04}, we studied the subtle stress tensor
correlations in non-vacuum states created by moving mirrors in two-dimensional
flat spacetime.
One of the key results of the present 
paper will be the derivation of a covariant
expression for the correlation function as a sum of total derivative terms.
This expression will be given in Sect.~\ref{sec:2Dcorr} for two dimensions
and in Sect.~\ref{sec:4Dcorr} for four dimensions, with the details of the
derivations presented in Appendices~\ref{sec:AppA} and  \ref{sec:AppB},
respectively. We will discuss spacetime averages of the energy density
correlation function in Sects.~\ref{sec:STaverage_2D} and 
\ref{sec:STaverage_4D}, and averages along a worldline in 
Sects.~\ref{sec:wl_sampling-2D} and \ref{sec:wl_sampling-4D}. The results
will be summarized and discussed in Sect.~\ref{sec:sum}. Units in which
$\hbar = c =1$, and a spacelike metric signature will be used throughout 
this paper.

\section{Two Dimensions}

\subsection{Covariant Stress Tensor Correlation Function}
\label{sec:2Dcorr}

We will be concerned with the stress tensor correlation function
\begin{equation}
C_{\mu \nu \alpha \beta}(x,x') = \langle :T_{\mu \nu}(x):\;
:T_{\alpha \beta}(x'): \rangle
\end{equation} 
for a massless, minimally coupled scalar field in two-dimensional
Minkowski spacetime in the vacuum state. Here $:T_{\mu \nu}(x):$
is the normal ordered stress tensor operator, so 
$\langle :T_{\mu \nu}(x): \rangle = 0$. We especially seek an expression
for $C_{\mu \nu \alpha \beta}(x,x')$ as a sum of terms, each of which
is a total derivative of a function with at most logarithmic singularities
as $x' \rightarrow x$. This will allow us to define integrals of the
correlation function by integration by parts. 

Such a form is derived in Appendix~\ref{sec:AppA}, where it is shown that
\begin{eqnarray}
C_{\mu \nu \alpha \beta}(x,x') &=& \frac{1}{384\, \pi^2}\; \biggl[
-8\, \partial_\mu\, \partial_\nu\,\partial_\alpha\,\partial_\beta\, f_1
 - 2\, g_{\mu\nu}\, g_{\alpha\beta}\, \Box \Box f_2 \nonumber \\
&+&  \, (g_{\mu\alpha}\, g_{\nu\beta} + 
g_{\mu\beta}\, g_{\nu\alpha})\, \Box \Box f_2  
+ 2\, (g_{\alpha\beta}\, \partial_\mu\, \partial_\nu\, +
 g_{\mu\nu}\,\partial_\alpha\,\partial_\beta) \Box f_2 \nonumber \\
&-& \,  (g_{\alpha\nu}\, \partial_\mu\, \partial_\beta\, +
g_{\alpha\mu}\, \partial_\nu\, \partial_\beta\, +
g_{\beta\nu}\, \partial_\mu\, \partial_\alpha\, +
g_{\beta\mu}\, \partial_\nu\, \partial_\alpha\ )\, \Box f_2 \biggl] \,, 
\label{eq:2Dcorr}     
\end{eqnarray}
where
\begin{equation}
f_1 = \ln(\Delta x^2/ \ell^2)\,,  
\end{equation}
and
\begin{equation}
f_2 = \ln^2(\Delta x^2/ \ell^2)\,, \label{eq:def_f2}
\end{equation} 
where $\ell$ is an arbitrary constant with dimensions of length. 
The correlation function is independent of the choice of $\ell$.
Here $\Box = \partial^\mu \, \partial_\mu$ is the wave operator, and
$\Delta x^2 = (x^\mu - {x'}^\mu)(x_\mu - {x'}_\mu)$. Because 
$\partial_\mu\,f_1 = \partial f_1/\partial x^\mu =  
 - \partial_{\mu'}\, f_1 = \partial f_1/\partial {x'}^\mu$, 
the correlation function,
Eq.~(\ref{eq:2Dcorr}), can be written in several equivalent forms.

The energy density correlation function becomes
\begin{equation}
C(x,x') = C_{ttt't'} =  -\frac{1}{48 \pi^2} \, \partial^4_t \,f_1
= -\frac{1}{48 \pi^2} \, \partial^2_t \partial^2_{t'}\,f_1 \,. \label{eq:2d-C}
\end{equation} 
Note that none of the $f_2$ terms contribute  in this case.
This expression allows us to compute the mean squared average energy density.
Let $g(t)$ be a time sampling function, and $h(x)$ be a spatial sampling
function. Then we define the averaged energy density operator as
\begin{equation}
\bar{\rho} = \int dt\, g(t)\,\int dx\, h(x)\, :T_{tt}: \,.
\end{equation} 
The mean square of this operator is
\begin{equation}
\hat{C} = \langle {\bar{\rho}}^2 \rangle =  
\int dt\, g(t)\,\int dx\, h(x)\,
 \int dt'\, g(t')\,\int dx'\, h(x')\, C(x,x') \,.  \label{eq:2d_hatC0} 
\end{equation} 
If we insert Eq.~(\ref{eq:2d-C}) into the above expression, and then
integrate by parts, we can write
\begin{equation}
\hat{C} = -\frac{1}{48 \pi^2} \,
\int dt\, \ddot{g}(t)\,\int dt'\, \ddot{g}(t')\,
\int dx\, h(x)\,\int dx'\, h(x')\, f_1 \,.
\end{equation} 
In the limit that the width of the spatial sampling function goes to zero,
$h(x) \rightarrow \delta(x)$ and we obtain
\begin{equation}
\hat{C} = -\frac{1}{48 \pi^2} \,
\int dt\, \ddot{g}(t)\,\int dt'\, \ddot{g}(t')\,\ln[(\Delta t)^2/\ell^2]\,.
     \label{eq:2d_hatC}
\end{equation}

\subsection{Averaging over Space and Time - 2D} 
\label{sec:STaverage_2D}

Rather than using Eq.~(\ref{eq:2d_hatC}), in some cases we can also directly
evaluate the integral in Eq.~(\ref{eq:2d_hatC0}) using contour integration
methods. For the explicit examples to be treated in this paper, the latter
approach is more convenient.   
The energy density correlation function, Eq.~(\ref{eq:2d-C}), can be expressed 
as  
\begin{equation} 
C(x,x') = \frac{{({\Delta t}^2 + {\Delta x}^2)}^2 +  
4 \, {\Delta t}^2 {\Delta x}^2}{4 \pi^2 \, 
{({\Delta t}^2 - {\Delta x}^2)}^4} \,, 
\label{eq:Cxx'} 
\end{equation} 
where $\Delta t = t-t'$, and $\Delta x= x-x'$.  
In this subsection we will sample this correlation function in 
both space and time  
with Lorentzian functions of width $\alpha$ in $t$ and $t'$,  
and $\beta$ in $x$ and $x'$. Further, let the spatial sampling functions  
coincide, but let the temporal ones be displaced by $t_0$.  
 
Let  
\begin{equation} 
\hat C(t_0) = \int_{-\infty}^{\infty} \, dt \, g_L(\alpha, t+t_0)  
\int_{-\infty}^{\infty} \, dt' \, g_L(\alpha, t')  
\int_{-\infty}^{\infty} \, dx \, g_L(\beta, x)  
\int_{-\infty}^{\infty} \, dx' \, g_L(\beta, x') \, C(x,x') \, 
\end{equation} 
where
\begin{equation} 
g_L(\alpha, t) = \frac{\alpha}{\pi(t^2+\alpha^2)} \, ,
\end{equation} 
and
\begin{equation}
\int_{-\infty}^{\infty} \, dt \, g_L(\alpha, t) = 1 \,.
\end{equation}
Now let $t \rightarrow t-t_0$, so that we have  
\begin{eqnarray} 
\hat C(t_0) &=& \int_{-\infty}^{\infty} \, dt \, g_L(\alpha, t) \, 
\int_{-\infty}^{\infty} \, dt' \, g_L(\alpha, t') \, 
\int_{-\infty}^{\infty} \, dx \, g_L(\beta, x) \,  \nonumber \\ 
&\,\,\,\,\,\,\,&\times 
\int_{-\infty}^{\infty} \, dx' \, g_L(\beta, x') \,\, C(t-t'-t_0,x-x')  
\nonumber \\ 
&=&\int_{-\infty}^{\infty} \, d\tau \,\, g_L(a, \tau) \, 
\int_{-\infty}^{\infty} \, d\rho \,\, g_L(b, \rho) \,\, 
C(\tau-t_0,\rho) \,, 
\end{eqnarray} 
where $a=2 \alpha,\,b=2 \beta,\,\tau=t-t'$ and $\rho=x-x'$. 
In the last step, we used the identity
\begin{equation}
\int_{-\infty}^{\infty} \, dt \, g_L(\alpha, t) \,
\int_{-\infty}^{\infty} \, dt' \, g_L(\alpha, t') \, F(t-t')
= \int_{-\infty}^{\infty} \, d\tau \,\, g_L(a, \tau) \, F(\tau)\,.
\end{equation} 

\begin{figure} 
\begin{center} 
\leavevmode\epsfysize=8cm\epsffile{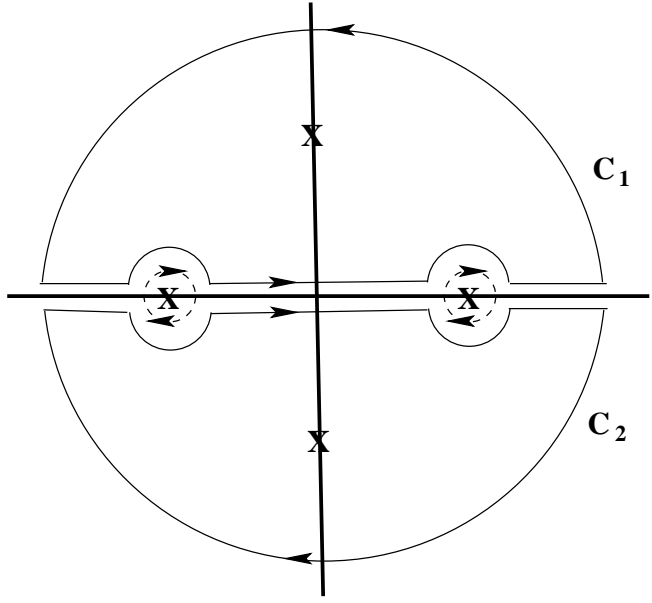} 
\end{center} 
\caption{Some possible integration contours for Eq.~(\ref{eq:2D_residues})
are illustrated. There are two simple poles on the imaginary axis, and two
higher order poles on the real axis. Both types of poles are denoted by the 
letter {\bf X}. The contours $C_1$ and $C_2$ both yield 
the same result for the integral. Integration around either of the poles
on the real axis (dashed line circles) give an imaginary result, so the real
part of the integral is independent of whether these poles are enclosed or
not. } 
\label{fig:contours} 
\end{figure}

We may do the integral on $\rho$ first, by contour integration. The  
integrand has simple poles at $\rho=\pm i b$ and fourth order poles at  
$\rho = \pm (\tau-t_0)$. We choose a contour in the upper half-plane  
which avoids the fourth order poles, the contour $C_1$ in 
Fig.~\ref{fig:contours}. In fact, we could use other contours such as
$C_2$ and still obtain the same answer. Even if we chose a contour which
enclosed either of the fourth order poles, our answer for the real part of 
the integral would still be the same. This is because the contribution of
either of these poles, the result of integrating around the closed circular
paths, is pure imaginary. Note that the straight segments and the
semicircular segments of  $C_1$ each contain real terms which diverge as the 
radii of the semicircles go to zero. However, these terms cancel when the 
straight and semicircular contributions are added. The divergent terms
on the straight segments are the boundary terms that would arise from
integrating by parts along these segments only. Thus integration by parts 
along the straight segments and discarding the boundary terms produces
the same  result as integration along the complete contour.

In any case, using the residue theorem we obtain  
\begin{equation} 
\int_{-\infty}^{\infty} d\rho \,\, g_L(b,\rho) \,\, C(\tau-t_0,\rho) = 
 \frac{{[{(\tau-t_0)}^2-b^2]}^2 - 4 b^2 \,{(\tau-t_0)}^2} 
{4 \pi^2 \,{[{(\tau-t_0)}^2+b^2]}^4} \,. \label{eq:2D_residues} 
\end{equation} 
The subsequent $\tau$-integration was performed and yields 
\begin{equation} 
\hat C (t_0) = \frac{[t_0(t_0+2a+2b)-{(a+b)}^2][t_0(t_0-2a-2b)-{(a+b)}^2]} 
{4 \pi^2 \,{[{t_0}^2+{(a+b)}^2]}^4} \,. 
\label{eq:hatC(t_0)} 
\end{equation}
(This and several other calculations in this paper were done using the public
domain algebraic manipulation program MAXIMA.)  
In the special case when $t_0 =0$, we simply have  
\begin{equation} 
\hat C(0) = \frac{1}{4 \pi^2 \, {(a+b)}^4} \,. 
\end{equation} 

Let us define 
\begin{equation} 
K(t_0,a,b) = \frac{\hat C(t_0)}{\hat C(0)} \,. \label{eq:def_K} 
\end{equation} 
In general, we have that $\hat C(t_0) = \hat C(-t_0)$. From 
Eqs.~(\ref{eq:hatC(t_0)}) and (\ref{eq:def_K}), we find that  
\begin{equation} 
\int_{-\infty}^{\infty} K(t_0,a,b) \, dt_0 = 0 \,,  
\end{equation} 
and similarly  
\begin{equation} 
\int_{0}^{\infty} K(t_0,a,b) \, dt_0 = 0 \,.  \label{eq:int_K}  
\end{equation} 
This result tells us that positively correlated regions ($K >0$), and 
anticorrelated regions  ($K <0$) have equal weight.

\subsection{Sampling along a Worldline - 2D} 
\label{sec:wl_sampling-2D} 

\begin{figure} 
\begin{center} 
\leavevmode\epsfysize=
8cm\epsffile{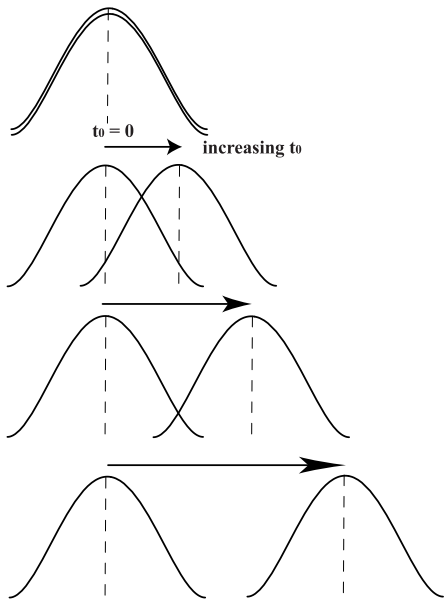} 
\end{center} 
\caption{Two sampling functions which initially coincide and are  
then gradually separated.} 
\label{fig:overlapping_sampling_functions-sketch} 
\end{figure}

In this subsection, we shall specialize to the case of sampling  
along a worldline, i.e., we will effectively set the width of  
the spatial sampling function to zero. 
Define a normal-ordered smeared stress tensor operator by  
\begin{equation} 
S(t_0)=\int_{-\infty}^{\infty} dt \, g(t,t_0) \, :T_{tt}(t):  \,, 
\label{eq:St_0} 
\end{equation} 
where $g(t,t_0)$ is a sampling function whose peak is at  $t = t_0$. 
Although $\langle S \rangle =0$ in the vacuum  
state, $\langle S^2 \rangle \neq 0$. From Eq.~(\ref{eq:2d_hatC}), 
we have that  
\begin{eqnarray} 
\langle S^2 \rangle &=& \int_{-\infty}^{\infty} dt \,  
\int_{-\infty}^{\infty} \, dt' \, g(t,t_0) \, g(t',t_0)  
\, C(t,t')  \nonumber \\ 
&=&-\frac{1}{48 \pi^2} \,\int_{-\infty}^{\infty} \, dt \, {\ddot g(t,t_0)} \, 
\int_{-\infty}^{\infty} \, dt' \, {\ddot g(t',t_0)} \, 
\ln[{\Delta t}^2/ \ell^2] \,. 
\end{eqnarray}  
The case we want to consider is two regions of time-sampled  
energy density which are allowed to initially coincide but  
which are then gradually separated from one another. One sampling function  
has its peak at $t'=0$ and the other at $t=t_0$. We want to imagine  
sliding these regions away from one another  
(see Fig.~\ref{fig:overlapping_sampling_functions-sketch}),  
and examine the behavior of the vacuum correlation function as we vary $t_0$.  
 
With Eq.~(\ref{eq:St_0}), we can write  
\begin{eqnarray} 
\langle S(t_0) S(0) \rangle &=& \int_{-\infty}^{\infty} dt \,  
\int_{-\infty}^{\infty} \, dt' \, g(t,t_0) \, g(t',0)  
\, C(t,t')  
\label{eq:St_0-S0_a}\\  
&=&-\frac{1}{48 \pi^2} \,\int_{-\infty}^{\infty} \, dt  
\,\int_{-\infty}^{\infty} \, dt' \, {\ddot g(t,t_0)} \, 
 {\ddot g(t',0)} \, 
\ln[{\Delta t}^2/ \ell^2] \,. 
\label{eq:St_0-S0} 
\end{eqnarray}  
This represents the smeared energy density correlation function for two  
displaced regions along a worldline. We can normalize this quantity by  
defining 
\begin{equation} 
K(t_0) = \frac{\langle S(t_0) S(0) \rangle}{\langle S^2(0) \rangle} \,. 
\label{eq:Kt_0} 
\end{equation}

\begin{figure} 
\begin{center} 
\leavevmode\epsfysize=10cm\epsffile{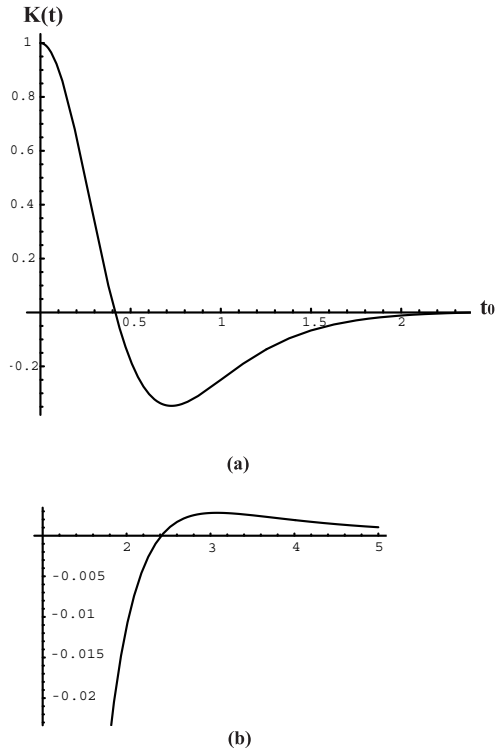} 
\end{center} 
\caption{The graph of $K(t_0)$ versus $t_0$ for a Lorentzian  
sampling function, in units with $a=1$. Here we have chosen $b=0$, so the 
sampling is in time only. Here (a) shows the overall form of $K(t_0)$, but
on a scale which does not reveal the final maximum. This peak is revealed
on a smaller scale graph, (b). } 
\label{fig:K-Lorentzian-2D_2} 
\end{figure}

As an example, we take the sampling function to be a Lorentzian. 
If we set $b=0$ and $a=1$ in Eq.~(\ref{eq:hatC(t_0)}), we  find  
\begin{equation} 
K(t_0) = \frac{(t_0^2-2t_0-1)(t_0^2+2t_0-1)}{{(t_0^2+1)}^4}  
=\frac{(1 - 6 t_0^2 + t_0^4)}{(1 + t_0^2)^4} \,. 
\end{equation}  
The choice of $b=0$ corresponds to sampling in time only, with displaced  
sampling functions. A plot of this function appears in  
Fig.~\ref{fig:K-Lorentzian-2D_2}(a). The plot is  
somewhat deceiving because there is actually a second positive peak which,  
on the scale of the plot, is too small to be seen. However, it must be there  
since $\hat C(t_0) \sim 1/(4 \, \pi^2 \, t_0^4)$, as 
$t_0 \rightarrow \infty$,  
and hence $K(t_0)$ has to approach $0$ from above for large $t_0$.  
The magnified view in Fig.~\ref{fig:K-Lorentzian-2D_2}(b) reveals the  
second positive peak. We can also see this by computing the extrema of 
$K(t_0)$ using 
\begin{equation} 
K'(t_0)= -\frac{4 t_0(t_0^4-10t_0^2+5)}{{(t_0^2+1)}^5} \,. 
\end{equation} 
One finds that $K'(t_0)=0$ at: $t_0 = 0$ (first maximum),  
$t_0 \approx 0.73$  
(minimum), and  $t_0 \approx 3.1$ (second maximum).  
 
\begin{figure} 
\begin{center} 
\leavevmode\epsfysize=8cm\epsffile{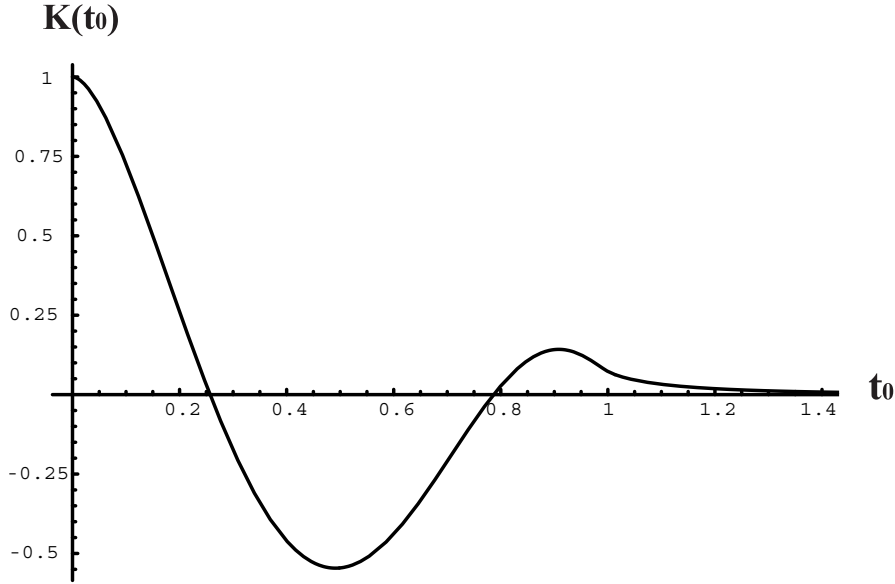} 
\end{center} 
\caption{The graph of $K(t_0)$ versus $t_0$ for the compactly supported  
sampling function given by Eq.~(\ref{eq:compact_f(t,t_0)}), in units  
where $a=1$.} 
\label{fig:K-compact-2D_2} 
\end{figure}

As a second example, consider a compactly supported sampling function of
width $a$ with $g=\dot g =0$ at $t= t_0 \pm a/2$ . 
A simple choice of function which has this form is  
\begin{equation} 
g(t,t_0)=g(t-t_0)= \frac{30}{a^5} {(t-t_0-a/2)}^2 \, {(t-t_0+a/2)}^2 \,. 
\label{eq:compact_f(t,t_0)} 
\end{equation} 
The second derivative of this function is  
\begin{equation} 
\ddot g(t-t_0) = \frac{30 [12{(t-t_0)}^2-a^2]}{a^5} \,, 
\label{eq:2nd_deriv_f} 
\end{equation} 
and 
\begin{equation} 
\langle S^2(0) \rangle = \frac{25}{2 \pi^2 a^4} \,. 
\label{eq:S^2_0} 
\end{equation} 
Using Eqs.~(\ref{eq:St_0-S0}), (\ref{eq:Kt_0}), (\ref{eq:2nd_deriv_f}),
and (\ref{eq:S^2_0}),  
one can evaluate $K(t_0)$, which is plotted as a function of $t_0$  
in Fig.~\ref{fig:K-compact-2D_2}. Note that the number of maxima and minima  
of $K(t_0)$ for the compactly supported sampling function, given in  
Eq.~(\ref{eq:compact_f(t,t_0)}), is the same as for the  
Lorentzian sampling function shown earlier. However, for  
the compactly supported sampling function case, the second  
maximum is much more pronounced. A calculation also shows that for 
both the Lorentzian and the compactly supported sampling functions, 
we have that   
\begin{equation} 
\int_0^\infty K(t_0) dt_0 = 0 \,. \label{eq:int_K0} 
\end{equation}

We will show in Appendix~\ref{sec:AppC} that this is true for  
arbitrary smooth sampling functions.  In this appendix, we also prove that
\begin{equation} 
 \langle S^2(0) \rangle > 0  \label{eq:S_sq} \,.
\end{equation}
 This establishes that the behavior 
illustrated in Fig.~\ref{fig:K-compact-2D_2} is independent of the details of 
the sampling function. 
The fact that $\langle S^2(0) \rangle > 0$ implies
that nearly overlapping regions are positively correlated with one another.
As $t_0$ increases, the correlation is replaced by anticorrelation, as shown 
by the negative minimum in $K(t_0)$. This anticorrelation implies that if 
we measure positive energy in a given region, there must be negative energy
found nearby. Finally, when the regions are sufficiently separated, the
positive correlation returns, as evidenced by the final positive peak in
Fig.~\ref{fig:K-compact-2D_2}. One can understand why disjoint regions
must be positively correlated from the fact that $C(x,x')  > 0$. When
$x \not= x'$ everywhere in the range of integration, then the integral for
$\hat{C}$ is well defined as an ordinary integral, and must be positive.
On the other hand, when we must integrate through  points where $x = x'$, 
then $C(x,x')$ becomes defined only as a distribution, and the integration
by parts procedure can produce a negative result.

\section{Four Dimensions}

\subsection{Covariant Stress Tensor Correlation Function }
\label{sec:4Dcorr}

In this section, we consider the vacuum stress tensor correlation function in
four dimensions. The covariant form of this function is derived in 
Appendix~\ref{sec:AppB}, with the result
\begin{eqnarray}
C_{\mu \nu \alpha \beta}(x,x') &=& -\frac{1}{61440\, \pi^4}\; \biggl[
 8\, \partial_\mu\, \partial_\nu\,
                  \partial_\alpha\,\partial_\beta\,\Box \Box f_2
+ 6\, g_{\mu\nu}\, g_{\alpha\beta}\,  \Box^4 f_2 \nonumber \\
&+&  \, (g_{\mu\alpha}\, g_{\nu\beta} + 
g_{\mu\beta}\, g_{\nu\alpha})\, \Box^4 f_2  
 -6\, (g_{\alpha\beta}\, \partial_\mu\, \partial_\nu\, +
 g_{\mu\nu}\,\partial_\alpha\,\partial_\beta)\, \Box^3 f_2 \nonumber \\
&-& \,  (g_{\alpha\nu}\, \partial_\mu\, \partial_\beta\, +
g_{\alpha\mu}\, \partial_\nu\, \partial_\beta\, +
g_{\beta\nu}\, \partial_\mu\, \partial_\alpha\, +
g_{\beta\mu}\, \partial_\nu\, \partial_\alpha\ )\, \Box^3 f_2 \biggl] \,.
\label{eq:4Dcorr}
\end{eqnarray}
Note that only the function $f_2$, defined in Eq.~(\ref{eq:def_f2}), 
appears here, in contrast to the two-dimensional result.
The energy density correlation function in four dimensions is given by
\begin{equation}
C(x,x') = C_{ttt't'} = - \frac{1}{7680 \pi^4}\, (\nabla^2)^2\,\Box^2 f_2 
=- \frac{1}{7680 \pi^4}\, \nabla^2\, {\nabla'}^2\, \Box\, \Box' f_2 \,,
                                         \label{eq:4D_C}
\end{equation} 
where $\nabla^2 = \Box + \partial^2_t$ is the three-dimensional 
Laplacian operator.
This form may be used to compute the  mean squared average energy density
over a spacetime region defined by a sampling function $F(x)$. If we define
\begin{equation}
\bar{\rho} = \int d^4x\, F(x)\, :T_{tt}: \,,
\end{equation} 
then
\begin{equation}
\hat{C} = \langle {\bar{\rho}}^2 \rangle =  \int d^4x\, F(x)\, 
\int d^4x'\, F(x')\, C(x,x') \,.
\end{equation} 
After an  integration by parts, this may be expressed as
\begin{equation}
\hat{C} = - \frac{1}{7680 \pi^4}\,  \int d^4x\,\nabla^2\, \Box F(x)\,
\int d^4x'\,{\nabla'}^2\, \Box' F(x')\, f_2(x-x') \,.
\end{equation}

At first sight, the process of obtaining finite spacetime averages of the 
correlation function may seem mysterious. We start with an expression for
$C_{\mu \nu \alpha \beta}(x,x')$ which diverges as $(x-x')^{-8}$ as
$x' \rightarrow x$, which seems to be a nonintegrable singularity. Yet
we nonetheless obtain finite integrals of this expression. The reason that
this is possible is that although $C_{\mu \nu \alpha \beta}(x,x')$ is 
singular as a function, it is a well-defined distribution. This is shown by 
the existence of the expression, Eq.~(\ref{eq:4Dcorr}), where
$C_{\mu \nu \alpha \beta}(x,x')$ is expressed as a sum of derivatives of
a function with no more than logarithmic singularities. An alternative
treatment of the singularities of stress tensor correlation functions
was given in Ref.~\cite{FW04}. There dimensional regularization was used
to render the correlation functions finite. In the limit in which 
$n \rightarrow 4$, where $n$ is the spacetime dimension, time-ordered
stress tensor correlation functions possess a pole term, which can be
absorbed in a renormalization  involving $R^2$ and $R_{\mu\nu}R^{\mu\nu}$
counterterms in the gravitational action. However, the correlation
functions without time ordering, such as $C_{\mu \nu \alpha \beta}(x,x')$,
have no pole term and are hence finite in dimensional regularization
in the $n \rightarrow 4$ limit. This is another way to understand why
$C_{\mu \nu \alpha \beta}(x,x')$ is a well-defined distribution, and why the
integration by parts method yields finite results.

\subsection{Averaging over Space and Time - 4D}
\label {sec:STaverage_4D}

Here we will perform a calculation analogous to that in 
Sect.~\ref{sec:STaverage_2D},
except involving averaging over space and time in four dimensions.
The energy density correlation function, Eq.~(\ref{eq:4D_C}), may be
expressed as 
\begin{equation}
C(x,x') = \frac{(\tau^2 +3r^2)(3\tau^2+r^2)}{2 \pi^4(\tau^2 -r^2)^6}
 \,,
                                         \label{eq:4D_C2}
\end{equation}
where $\tau = t -t'$ and $r = |\mathbf{x} - \mathbf{x'}|$.  
As before, we use Lorentzian sampling functions of width $\alpha$ in 
$t$ and in $t'$. The time-averaged correlation function is
\begin{equation}
\hat{C}_T = \int^\infty_{-\infty}\, dt \,  g_L(\alpha, t) \,
   \int^\infty_{-\infty}\, dt' \,  g_L(\alpha, t') \, C(x,x') 
= \int^\infty_{-\infty}\, d\tau \, g_L(a,\tau)\, C(x,x') \, ,
  \label{eq:4D_C3}
\end{equation} 
where $a= 2\,\alpha$. The integrand in the $\tau$ integral has first
order poles at $\tau = \pm i a$ and sixth order poles at $\tau = \pm r$.
The integral may be performed by contour integration in a way analogous to
the integral in Eq.~(\ref{eq:2D_residues}). The result is
\begin{equation}
\hat{C}_T = \frac{(3 r^2 -a^2)(r^2 -3 a^2)}{2 \pi^4(r^2 + a^2)^6} \,.
                                       \label{eq:C_T}
\end{equation} 

Next we wish to average $\hat{C}_T$ over the spatial directions. Here 
it will be convenient to use a Gaussian sampling function
\begin{equation}
g_G(\beta,x) = \frac{1}{\sqrt{\pi}\, \beta}\, {\rm e}^{-x^2/\beta^2} \,,
\end{equation} 
in each of the Cartesian space coordinates, $x,y,z,x',y',z'$, and define
the spacetime average as 
\begin{eqnarray}
\hat{C} &=&  \int^\infty_{-\infty}\, dx\, g_G(\beta,x)
  \int^\infty_{-\infty}\, dy\, g_G(\beta,y) 
\int^\infty_{-\infty}\, dz\, g_G(\beta,z) \nonumber \\
&\times& \int^\infty_{-\infty}\, dx'\, g_G(\beta,x')
  \int^\infty_{-\infty}\, dy'\, g_G(\beta,y') 
\int^\infty_{-\infty}\, dz'\, g_G(\beta,z') \, \hat{C}_T \,.
\end{eqnarray} 
We may use the fact that
\begin{equation}
 \int^\infty_{-\infty}\, dx\, g_G(\beta,x)
\int^\infty_{-\infty}\, dx'\, g_G(\beta,x')\, f(x-x')
=  \int^\infty_{-\infty}\, d \Delta x\, g_G(b,\Delta x) \,f(\Delta x) \,,
\end{equation} 
where $\Delta x = x - x'$ and $b = \sqrt{2}\, \beta$.
This leads to
\begin{eqnarray}
\hat{C} &=&  \int^\infty_{-\infty}\, d \Delta x\, g_G(b,\Delta x)
  \int^\infty_{-\infty}\, d \Delta y\, g_G(b,\Delta y) 
\int^\infty_{-\infty}\, d \Delta z\, g_G(b,\Delta z) \, \hat{C}_T \nonumber \\
&=& \frac{4}{\sqrt{\pi} b^3}\,  \int^\infty_0 dr \, r^2 \, 
 {\rm e}^{-r^2/\beta^2} \,  \hat{C}_T \,,
\end{eqnarray}
where $r^2 = (\Delta x)^2 + (\Delta y)^2 + (\Delta z)^2 $. 
If we use Eq.~(\ref{eq:C_T}), then we can write the spacetime averaged
correlation function as
\begin{equation}
\hat{C} = \frac{2}{\pi^{9/2}\, b^3} \,  \int^\infty_0 dr \, r^2 \,
\frac{(3 r^2 -a^2)(r^2 -3 a^2)}{(r^2 + a^2)^6}\; 
{\rm e}^{-r^2/\beta^2} \,.    \label{eq:4D_STav}
\end{equation} 
The integral in the above expression may evaluated in terms of the error
function, erf, as
\begin{eqnarray}
\hat{C} &=& \frac{1}{15 \pi^4 a\, b^{13}}\,\biggl\{ \sqrt{\pi}\,
\left[1 - {\rm erf}\left(\frac{a}{b}\right)\right] {\rm e}^{a^2/b^2}\,
(15b^6 +90a^2b^4+60a^4b^2+8a^6)  \nonumber \\
 &-& 2ab(3b^2+2a^2)(11b^2+2a^2) \biggr\} \,.
\label{eq:4D_STav2}
\end{eqnarray} 

Now we wish to discuss the limits in which one sampling length scale is small
compared to the other. First consider the case of a small spatial scale,
$b \ll a$. The exponential factor in Eq.~(\ref{eq:4D_STav}) guarantees
that only values of $r \alt b$ contribute. Thus we can assume that $r \ll a$
in the integrand and write
\begin{equation}
\frac{(3 r^2 -a^2)(r^2 -3 a^2)}{(r^2 + a^2)^6} \approx \frac{3}{a^8} \,.
\end{equation}
Then we have
\begin{equation}
\hat{C} \approx \frac{3}{2\, \pi^4\, a^8}  \label{eq:small_b}
\end{equation}
when $b \ll a$. This shows that only temporal sampling is necessary
in order for $\hat{C}$ to be finite. Equation~(\ref{eq:small_b}) may
also be derived from the explicit form, Eq.~(\ref{eq:4D_STav2}), by use
of the asymptotic form of the error function for large argument.

Next we consider the opposite limit, where $a\ll b$. However, 
$\hat{C} \rightarrow \infty$ as $a \rightarrow 0$ for fixed, nonzero $b$.
This may be seen from the integral, Eq.~(\ref{eq:4D_STav}), which becomes
proportional to $\int^\infty_0 dr \, r^{-6} \,  {\rm e}^{-r^2/\beta^2}$
as  $a \rightarrow 0$. Alternatively, we can expand Eq.~(\ref{eq:4D_STav2})
for small $a$ and show that
\begin{equation}
\hat{C} \sim \frac{1}{\pi^{7/2}\,a\,b^7}\,, 
\quad {\rm as}\quad a \rightarrow 0 \,.
\end{equation}
Thus in four dimensions, averaging over space alone is not
sufficient to lead to a finite mean squared energy density. This result
was obtained previously by Guth~\cite{Guth} and by Roura~\cite{Roura}.

\subsection{Sampling along a Worldline - 4D}
\label{sec:wl_sampling-4D}

In the previous subsection, we found that it is possible to take the limit
of a vanishing spatial sampling scale, so that one is sampling along a 
worldline. Here we will consider that limit for displaced temporal sampling 
functions. First consider Lorentzian sampling functions and let
\begin{equation}
\hat{C}(t_0,r) =  \int^\infty_{-\infty}\, dt \,  g_L(\alpha, t+t_0) \,
   \int^\infty_{-\infty}\, dt' \,  g_L(\alpha, t') \, C(\tau,r) 
= \int^\infty_{-\infty}\, d\tau \, g_L(a,\tau-t_0)\, C(\tau,r) \, ,
\end{equation}  
where $C(\tau,r)$ is given by Eq.~(\ref{eq:4D_C2}), and $a = 2\alpha$. 
If we were to sample in space with a function whose 
width is small compared to $a$, the result is the same as setting $r=0$ in 
the above expression. More precisely, we perform the integral for nonzero 
$r$, using the same method as used to obtain Eq.~(\ref{eq:2D_residues}), 
and then take the $r \rightarrow 0$ limit. The result is
\begin{equation}
\hat{C}(t_0,0) = 
\frac{(t_0^4-4a t_0^3-6a^2 t_0^2+4a^3 t_0 +a^4)  
      (t_0^4+4a t_0^3-6a^2 t_0^2-4a^3 t_0 +a^4)}{\pi^4 (t_0^2 +a^2)^8} \,.
                                     \label{eq:4D_dspl}
\end{equation}
This function has a form similar to that illustrated in 
Fig.~\ref{fig:K-Lorentzian-2D_2}, except that it has three maxima and two 
minima. It is somewhat difficult to graph because the relative sizes
of the extrema decrease very rapidly.

 In the limit that $r =0$,
we may write the four-dimensional correlation function as
\begin{equation}
C(t,t') = \frac{3}{2 \pi^4\, (t-t')^8} = 
-\frac{1}{6720 \pi^4}\, \partial_t^4\, \partial_{t'}^4\, \ln[(t-t')^2/\ell^2]
  \, .
\end{equation}
We can sample the energy density with arbitrary 
displaced sampling functions and write
\begin{eqnarray} 
\langle S(t_0) S(0) \rangle &=& \int_{-\infty}^{\infty} dt \,  
\int_{-\infty}^{\infty} \, dt' \, g(t-t_0) \, g(t')  
\, C(t,t')  \label{eq:4D_worldline} \\  
&=& -\frac{1}{6720 \pi^4} \,\int_{-\infty}^{\infty} 
\, dt  
\,\int_{-\infty}^{\infty} \, dt' \, [\partial_t^4\, g(t-t_0)] \, 
 [\partial_{t'}^4\, g(t')] \, 
\ln[{\Delta t}^2/ \ell^2] . 
\label{eq:4D_compact_disp} 
\end{eqnarray} 
It should be noted that here we did not use the form of the energy density
correlation function, Eq.~(\ref{eq:4D_C}), which follows from the 
covariant form.   Instead, we let $r \rightarrow 0$, and then expressed the
result in terms of time derivatives of a logarithmic function. A more
rigorous approach would be to  average Eq.~(\ref{eq:4D_C}) over both
space and time, and then let the widths of the spatial sampling functions 
go to zero. However, this
is difficult to do explicitly with general sampling functions.   
The equivalence of the two approaches needs to be
studied more carefully.

\begin{figure} 
\begin{center} 
\leavevmode\epsfysize=8cm\epsffile{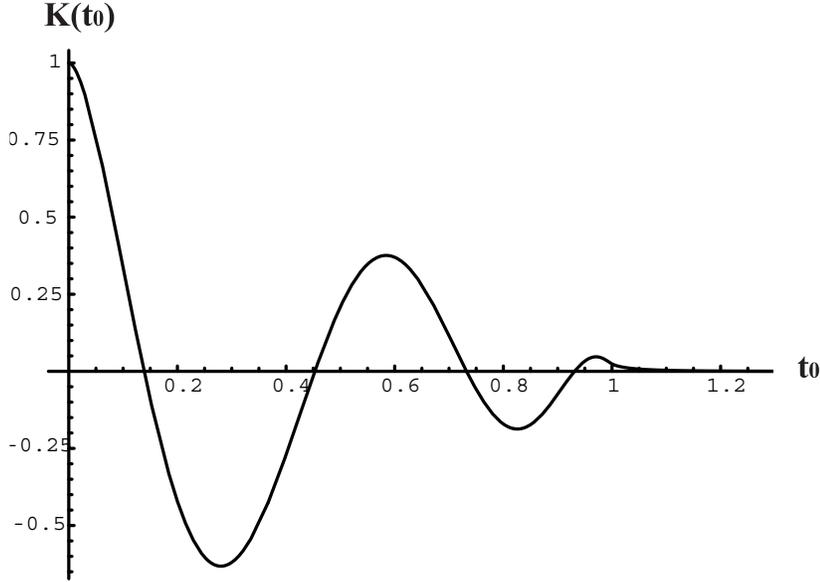} 
\end{center} 
\caption{The graph of $K(t_0)$  in four dimensions as a function of $t_0$ 
for the compactly supported  
sampling function given by Eq.~(\ref{eq:4D_compact}), in units  
where $a=1$.} 
\label{fig:4D_K0} 
\end{figure}

Let us next consider a compactly supported sampling function given by
\begin{equation}
g(t) = \frac{630}{a^9}\,(t-  a/2)^4\, (t +  a/2)^4 \, ,
      \label{eq:4D_compact}
\end{equation}
for $|t|\leq a/2$, and $g(t) = 0$ for $|t|\geq a/2$. 
Note that $g(t)$ and its first three derivatives vanish at 
$t = \pm \frac{1}{2} a$, so all surface terms vanish when we integrated by 
parts in Eq.~(\ref{eq:4D_compact_disp}) 
to obtain the second form for $\langle S(t_0) S(0) \rangle$. We may
again define $K(t_0)$ by Eq.~(\ref{eq:Kt_0}) and evaluate it numerically.
The result is plotted in Fig.~\ref{fig:4D_K0}. 

As in two dimensions, there are regions of correlation and of anticorrelation
as $t_0$ increases. However, the behavior in four dimensions is more 
complicated, with three maxima and two minima. This appears to be due to
the greater number of derivatives of the sampling function in
Eq.~(\ref{eq:4D_compact_disp}), as compared to Eq.~(\ref{eq:St_0-S0}). 

In Appendix \ref{sec:AppC}, we show that in two and four dimensions
\begin{equation} 
\int_{0}^{\infty} K(t_0) dt_0 = 0 \,,  
\end{equation} 
 and that
\begin{equation}
\langle S^2(0) \rangle > 0 \, ,
\end{equation}
for a general $g(t)$.
From Eq.~(\ref{eq:4D_dspl}), we can also explicitly verify that
$\int_{0}^{\infty} K(t_0) dt_0 = 0$ for the Lorentzian sampling
function.

\section{Summary}
\label{sec:sum}

In this paper, we have presented covariant expressions for the Minkowski vacuum
stress tensor correlation function in two dimensions, Eq.~(\ref{eq:2Dcorr}),
and in four dimensions, Eq.~(\ref{eq:4Dcorr}). These expressions are of the
form of a sum of terms, each of which is a derivative of a scalar function
with logarithmic singularities in the coincidence limit. These expressions
allow one to write spacetime averages of the correlation function as finite
integrals. We explicitly evaluated such averages of the energy density 
in two dimensions using Lorentzian sampling functions in both space and time.
The resulting expression, Eq.~(\ref{eq:hatC(t_0)}), is symmetric in the
spatial and temporal sampling widths, and is finite as either width goes
to zero with the other width fixed at a nonzero value.  

We next studied
the correlations of the sampled 2D energy density along a worldline
using displaced sampling functions. This reveals the correlation and     
anticorrelation of measurements of the energy density  in overlapping
intervals. The result is illustrated in Fig.~\ref{fig:K-compact-2D_2} for
a compactly supported sampling function. 
When the intervals nearly overlap, the two 
measurements are positively correlated, as expected. When the overlap
has decreased somewhat, the two measurements become anticorrelated. This
can be interpreted as telling us that if we find energy density of one sign 
on the first measurement, we should find the opposite sign on the next
measurement. Finally, as the intervals become disjoint, the measurements
are again positively correlated. Furthermore, we show that for an arbitrary
sampling function, the net area under the correlation graph, e.g., the
one depicted in Fig.~\ref{fig:K-compact-2D_2}, is equal to zero. It is
hoped that further investigation will elucidate this interesting behavior.

The analogous calculation in four dimensions yields similar results.
However, in this case there are  two regions of anticorrelation and three of
positive correlation. The fluctuations in the averaged energy density
remain finite in the limit that the spatial width vanishes, but not in 
the limit that the temporal width goes to zero. Thus in four dimensions, the 
averaged energy density correlation function requires averaging in time
to be finite. 

There is a vaguely analogous result concerning quantum inequalities on the
averaged expectation value of the stress tensor in an arbitrary state.
There are finite lower bounds on the expectation value of the energy density
averaged on a worldline in both 2D and 4D, and on the spatial average in
2D. However, the spatial average in 4D has no lower bound~\cite{FHR}. 
The search for a deeper link between quantum inequalities and the vacuum
stress tensor correlation function is a topic for future research.

Another question which needs to be explored further is that of the 
physical effects of the passive metric fluctuations driven by
vacuum stress tensor fluctuations. One approach is that adopted in
Ref.~\cite{BF} where the Raychaudhuri equation was used as a Langevin
equation to study the luminosity fluctuations and angular blurring of
a distant source produced by passive metric fluctuations. The case
of the Minkowski vacuum was briefly discussed in Ref.~\cite{BF}, where it
was found that the natural quantum uncertainty in the test particles
used to probe the fluctuating geometry tends to hide the effects
of the metric fluctuations. However, this does not necessarily mean that these 
fluctuations are in principle unobservable. This is another question
for further study.

\begin{acknowledgments}
 We would like to thank Albert Roura for valuable comments and Alan Guth for
providing his unpublished notes.
 This work was supported in part by the National
Science Foundation under Grants PHY-0139969 and PHY-0244898.
\end{acknowledgments}

\appendix
\section{}
\label{sec:AppA}

In this appendix, we give the derivation of the stress tensor correlation 
function in two-dimensional spacetime in the Minkowski vacuum state. 
We first start with the form of the stress tensor for a massless, minimally
coupled scalar field:
\begin{equation}
T_{\mu \nu} = \phi_{,\mu} \phi_{,\nu} - \frac{1}{2} g_{\mu\nu}\,
                \phi^{,\rho} \phi_{,\rho} \,.
\end{equation}
From this expression we find the correlation function
\begin{eqnarray}
C_{\mu \nu \alpha \beta}(x,x') &=& \langle :T_{\mu \nu}(x):\;
:T_{\alpha \beta}(x'): \rangle =
\langle : \partial_\mu \phi\, \partial_\nu \phi: \;
 :\partial_{\alpha'} \phi \partial_{\beta'} \phi: \rangle \nonumber \\
&-&\frac{1}{2} g_{\mu\nu}\, \langle :\partial^\rho \phi\, \partial_\rho \phi:
   \; 
:\partial_{\alpha'} \phi \,\partial_{\beta'} \phi: \rangle  
- \frac{1}{2} g_{\alpha\beta}\,
      \langle : \partial_\mu \phi\,\partial_\nu \phi: \; 
:\partial^{\sigma'} \phi \,\partial_{\sigma'} \phi: \rangle \nonumber \\ 
&-& \frac{1}{4} g_{\mu\nu}g_{\alpha\beta}\, 
\langle : \partial^\rho \phi\,\partial_\rho \phi: \; 
:\partial^{\sigma'} \phi \,\partial_{\sigma'} \phi: \rangle \,. \label{eq:C1}
\end{eqnarray}
Here unprimed indices refer to the point $x$ and primed indices to $x'$.
Next we use the identity
\begin{equation}
\langle :\phi_1 \, \phi_2:\; :\phi_3 \, \phi_4: \rangle =
\langle \phi_1 \,\phi_3 \rangle \, \langle \phi_2 \,\phi_4 \rangle
+ \langle \phi_1 \,\phi_4 \rangle \, \langle \phi_2 \,\phi_3 \rangle\, ,
\end{equation}
where the $\phi_i$ are quantum fields or derivatives of quantum fields.
From this identity, we can show that
\begin{equation}
\langle : \partial_\mu \phi\, \partial_\nu \phi: \;
 :\partial_{\alpha'} \phi \partial_{\beta'} \phi: \rangle =
(\partial_\mu \, \partial_{\alpha'} \,D)
(\partial_\nu \, \partial_{\beta'} \,D) +
(\partial_\mu \, \partial_{\beta'} \,D)
(\partial_\nu \, \partial_{\alpha'} \,D)\, ,
 \label{eq:4phi} 
\end{equation} 
where
\begin{equation}
D = D(x,x') = \langle \phi(x) \phi(x') \rangle
\end{equation} 
is the two-point function. We can express the correlation function
in terms of derivatives of D as
\begin{eqnarray}
C_{\mu \nu \alpha \beta}(x,x') &=& 
(\partial_\mu \, \partial_{\alpha'} \,D)
(\partial_\nu \, \partial_{\beta'} \,D) +
(\partial_\mu \, \partial_{\beta'} \,D)(\partial_\nu \, \partial_{\alpha'} \,D)
                                              \nonumber \\
         &-&g_{\mu\nu}\,
(\partial^\rho\, \partial_{\alpha'} \,D)
(\partial_\rho\, \partial_{\beta'} \,D) 
  -  g_{\alpha\beta}\,
(\partial_\mu \, \partial^{\sigma'}\,D)(\partial_\nu \, \partial_{\sigma'}\,D)
                                           \nonumber \\
 &-&  \frac{1}{2} g_{\mu\nu}g_{\alpha\beta}\,
(\partial^\rho\, \partial^{\sigma'}\,D)(\partial_\rho\, \partial_{\sigma'}\,D)
 \, .  \label{eq:C2}
\end{eqnarray}
An equivalent expression for the case of a massive, nonminimal scalar field
has been given by Martin and Verdaguer. (See Eq.~3.42 in Ref.~\cite{MV99}.)
The analogous expression for the electromagnetic field is given in
Ref.~\cite{FW03}.

Up to this point, our treatment applies to spacetimes of any dimensionality.
Now we specialize to two-dimensional Minkowski spacetime. There is an
infrared divergence in the two-point function for a massless scalar field
in the Minkowski vacuum state in  two dimensions. Thus, the field must
either have a nonzero mass, or else the only physically allowed states
are ones which break Lorentz invariance~\cite{FV86}. Fortunately, the details
of either approach have no effect on our results. If we let the scalar
field have a small mass $m$, then the two-point function is given by
\begin{equation}
D = - \frac{1}{4\pi}\, \ln(c\, m^2\, \Delta x^2) 
\end{equation}  
in the limit that $ m^{-2} \gg \Delta x^2$. Here $c$ is a dimensionless
constant and $\Delta x^2 = (x^\mu - {x'}^\mu)(x_\mu - {x'}_\mu)$. 
Because the stress tensor correlation
function depends only upon derivatives of $D$, it is independent of $c$
and $m$. The second derivative of $D$ is
\begin{equation}
\partial_\mu \, \partial_{\alpha'} \,D = 
 -\frac{2 \Delta x_\mu\, \Delta x_\alpha - g_{\mu\alpha}\, \Delta x^2}
       {2 \pi\, (\Delta x^2)^2 } \, ,  \label{eq:partialD}
\end{equation} 
where $\Delta x_\mu = x_\mu - x'_\mu$. We can now combine Eqs.~(\ref{eq:C2})
 and (\ref{eq:partialD}) to obtain an explicit expression
for the stress tensor correlation function in two dimensions:
\begin{eqnarray}
C_{\mu \nu \alpha \beta}(x,x') &=& \frac{1}{4 \pi^2}\, \biggl[
\frac{8}{(\Delta x^2)^4} \; 
 \Delta x_\mu\ \Delta x_\nu \Delta x_\alpha \Delta x_\beta 
       \nonumber \\   
&-& \frac{2}{(\Delta x^2)^3} \; (g_{\mu\alpha}\,\Delta x_\nu\, \Delta x_\beta
+ g_{\mu\beta}\,\Delta x_\nu\, \Delta x_\alpha
+ g_{\nu\alpha}\,\Delta x_\mu\, \Delta x_\beta
+ g_{\nu\beta}\,\Delta x_\mu\, \Delta x_\alpha) 
       \nonumber \\   
&+& \frac{1}{(\Delta x^2)^2} \; (g_{\mu\alpha}\, g_{\nu\beta} +
g_{\mu\beta}\, g_{\nu\alpha} - g_{\mu\nu}\, g_{\alpha\beta}) \biggr]
                                  \label{eq:2DC1}
\end{eqnarray}

We next wish to express $C_{\mu \nu \alpha \beta}(x,x')$ as a sum of 
derivatives of scalar functions. Lorentz symmetry  suggests that these
be functions of $\Delta x^2$. Let $f = f(\Delta x^2)$. Then  the derivatives
of $f$ are
\begin{equation}
\partial_\mu\, f = 2\, \Delta x_\mu\, f' \,,
\end{equation} 
\begin{equation}
\partial_\mu\, \partial_\nu\, f = 2\,g_{\mu\nu}\,f' + 
4 \Delta x_\mu \Delta x_\nu \, f'' \,,
\end{equation} 
\begin{equation}
\partial_\mu\, \partial_\nu\,\partial_\alpha\, f = 
4 (g_{\mu\nu}\,\Delta x_\alpha + g_{\mu\alpha}\,\Delta x_\nu 
+ g_{\nu\alpha}\,\Delta x_\mu) f'' 
+8 \Delta x_\mu\ \Delta x_\nu \Delta x_\alpha \, f''' \, ,
\end{equation}
and 
\begin{eqnarray}
\partial_\mu\, \partial_\nu\,\partial_\alpha\,\partial_\beta\, f &=&
4(g_{\mu\alpha}\, g_{\nu\beta} + g_{\mu\beta}\, g_{\nu\alpha} + 
g_{\mu\nu}\, g_{\alpha\beta}) f''
+ 8 ( g_{\mu\alpha}\,\Delta x_\nu\, \Delta x_\beta
+ g_{\mu\beta}\,\Delta x_\nu\, \Delta x_\alpha \nonumber \\
&+& g_{\nu\alpha}\,\Delta x_\mu\, \Delta x_\beta
+ g_{\nu\beta}\,\Delta x_\mu\, \Delta x_\alpha
+ g_{\mu\nu}\,\Delta x_\alpha \Delta x_\beta
+ g_{\alpha\beta}\, \Delta x_\mu\,\Delta x_\nu ) \, f''' \nonumber \\
&+& 16 \Delta x_\mu\ \Delta x_\nu \Delta x_\alpha \Delta x_\beta\, f'''' \,. 
\end{eqnarray}
Here primes denote derivatives of $f$ with respect to its argument.
We will also need some expressions involving the wave operator:
\begin{equation}
\Box f = \partial_\mu \partial^\mu f = 2 n f' +4 \Delta x^2 f'' \,,
\label{eq:1box} 
\end{equation} 
\begin{equation}
\Box \Box f = 4n(n+2) f'' + 16(n+2) \Delta x^2 f''' 
    + 16 ( \Delta x^2)^2 f''''  \, , \label{eq:2box}
\end{equation}
and 
\begin{eqnarray}
\partial_\mu\, \partial_\nu\,\Box f &=& 4(n+2) g_{\mu\nu}\,f'' 
+8 [g_{\mu\nu}\,\Delta x^2 + (n+4) \Delta x_\mu\ \Delta x_\nu]\, f''' 
        \nonumber \\
&+& 16 \Delta x_\mu\ \Delta x_\nu \,\Delta x^2 f''' \, ,   
\end{eqnarray} 
where $n$ is the dimension of the spacetime.

Our goal is to express  $C_{\mu \nu \alpha \beta}(x,x')$ as a sum of 
derivatives acting on one or more choices of $f$.  Because 
$\partial_\mu\,f_1 = - \partial_{\mu'}\, f_1$, we can write our results in
several equivalent forms, but here and in Appendix \ref{sec:AppB} we will use 
derivatives with unprimed indices. If $f$ is dimensionless,
then in two dimensions we will need four derivatives in each term  
in order that
$C_{\mu \nu \alpha \beta}(x,x')$ has dimensions of ${\rm length}^{-4}$.
There are five fourth-rank tensors that we can form which have the
correct dimensions and symmetry properties:
\begin{equation}
\partial_\mu\, \partial_\nu\,\partial_\alpha\,\partial_\beta\, f \,,
\end{equation} 
 \begin{equation}
(g_{\alpha\beta}\, \partial_\mu\, \partial_\nu\, +
 g_{\mu\nu}\,\partial_\alpha\,\partial_\beta) \Box f \, ,
\end{equation} 
 \begin{equation}
 (g_{\alpha\nu}\, \partial_\mu\, \partial_\beta\, +
g_{\alpha\mu}\, \partial_\nu\, \partial_\beta\, +
g_{\beta\nu}\, \partial_\mu\, \partial_\alpha\, +
g_{\beta\mu}\, \partial_\nu\, \partial_\alpha\ ) \Box f \, ,
\end{equation} 
\begin{equation}
g_{\mu\nu} \,g_{\alpha\beta}\, \Box \Box f \,,
\end{equation}
and
\begin{equation}
(g_{\mu\alpha}\, g_{\nu\beta} + g_{\mu\beta}\, g_{\nu\alpha})\Box \Box f \,.
\end{equation}

We would like $f$ to have an integrable singularity at $\Delta x^2 = 0$,
so a natural choice is a power of a logarithmic function. First consider
\begin{equation}
f_1 = \ln(\Delta x^2/ \ell^2)\,,  \label{eq:f1def}
\end{equation} 
where $\ell$ is an arbitrary constant with dimensions of length.
However, $\Box f_1 =0$ in two dimensions, so the only nonzero tensor  
from the above list which can be formed from $f_1$ is
\begin{eqnarray}
\partial_\mu\, \partial_\nu\,\partial_\alpha\,\partial_\beta\, f_1 &=&
-\frac{96}{(\Delta x^2)^4} \; 
 \Delta x_\mu\ \Delta x_\nu \Delta x_\alpha \Delta x_\beta 
       \nonumber \\   
&+& \frac{16}{(\Delta x^2)^3} \; (g_{\mu\alpha}\,\Delta x_\nu\, \Delta x_\beta
+ g_{\mu\beta}\,\Delta x_\nu\, \Delta x_\alpha
+ g_{\nu\alpha}\,\Delta x_\mu\, \Delta x_\beta   \nonumber \\
&+& g_{\nu\beta}\,\Delta x_\mu\, \Delta x_\alpha
+ g_{\mu\nu}\,\Delta x_\alpha \Delta x_\beta
+ g_{\alpha\beta}\, \Delta x_\mu\,\Delta x_\nu ) 
       \nonumber \\   
&-& \frac{4}{(\Delta x^2)^2} \; (g_{\mu\alpha}\, g_{\nu\beta} +
g_{\mu\beta}\, g_{\nu\alpha} + g_{\mu\nu}\, g_{\alpha\beta}) \,.  \label{eq:f1}
\end{eqnarray}
This is not sufficient to form $C_{\mu \nu \alpha \beta}(x,x')$, so we
need another choice of $f$, which we take to be
\begin{equation}
f_2 = \ln^2(\Delta x^2/ \ell^2)\,.
\end{equation} 
From Eqs.~(\ref{eq:1box}) and (\ref{eq:2box}) with $n = 2$, we find
\begin{equation}
\Box f_2 = \frac{8}{\Delta x^2}
\end{equation}
and 
\begin{equation}
\Box \Box f_2 = \frac{32}{(\Delta x^2)^2} \,.
\end{equation} 
This allows us to form four tensors from $f_2$ with the correct symmetry
properties and dimension:
\begin{equation}
g_{\mu\nu}\, g_{\alpha\beta}\, \Box \Box f_2 = 
32 \,g_{\mu\nu}\, g_{\alpha\beta}\, \frac{1}{(\Delta x^2)^2} \, ,
                                                   \label{eq:t1} 
\end{equation} 
\begin{equation}
(g_{\mu\alpha}\, g_{\nu\beta} + g_{\mu\beta}\, g_{\nu\alpha})\, \Box \Box f_2 =
32\, (g_{\mu\alpha}\, g_{\nu\beta} + g_{\mu\beta}\, g_{\nu\alpha})\,
\frac{1}{(\Delta x^2)^2} \, , \label{eq:t2}
\end{equation} 
\begin{equation}
(g_{\alpha\beta}\, \partial_\mu\, \partial_\nu\, +
 g_{\mu\nu}\,\partial_\alpha\,\partial_\beta) \Box f_2 =
-32\,\frac{g_{\mu\nu}\, g_{\alpha\beta}}{(\Delta x^2)^2} 
+64\,\, \frac{ g_{\mu\nu}\,\Delta x_\alpha\, \Delta x_\beta +
g_{\alpha\beta}\, \Delta x_\mu\ \Delta x_\nu}{(\Delta x^2)^3} \, ,\label{eq:t3}
\end{equation} 
and 
\begin{eqnarray}
& &   (g_{\alpha\nu}\, \partial_\mu\, \partial_\beta\, +
g_{\alpha\mu}\, \partial_\nu\, \partial_\beta\, +
g_{\beta\nu}\, \partial_\mu\, \partial_\alpha\, +
g_{\beta\mu}\, \partial_\nu\, \partial_\alpha\ ) \Box f_2 =
-32\,\, \frac{g_{\mu\alpha}\, g_{\nu\beta} +g_{\mu\beta}\, g_{\nu\alpha}}
{(\Delta x^2)^2}   \nonumber \\
&+& \frac{64}{(\Delta x^2)^3}\, (g_{\mu\alpha}\,\Delta x_\nu\, \Delta x_\beta
+ g_{\mu\beta}\,\Delta x_\nu\, \Delta x_\alpha
+ g_{\nu\alpha}\,\Delta x_\mu\, \Delta x_\beta
+ g_{\nu\beta}\,\Delta x_\mu\, \Delta x_\alpha)  \,. \label{eq:t4} 
\end{eqnarray}
Note that 
$\partial_\mu\, \partial_\nu\,\partial_\alpha\,\partial_\beta\, f_2$  
is not a suitable term because it contains logarithmic pieces that do not
appear in $C_{\mu \nu \alpha \beta}(x,x')$ and which cannot be cancelled by 
any other terms. This leaves us with five tensors from which to form
the stress tensor correlation function. 

Let
\begin{eqnarray}
C_{\mu \nu \alpha \beta}(x,x') &=& \frac{1}{384\, \pi^2}\; \biggl[
c_1\, \partial_\mu\, \partial_\nu\,\partial_\alpha\,\partial_\beta\, f_1
+ c_2\, g_{\mu\nu}\, g_{\alpha\beta}\, \Box \Box f_2 \nonumber \\
&+& c_3 \, (g_{\mu\alpha}\, g_{\nu\beta} + 
g_{\mu\beta}\, g_{\nu\alpha})\, \Box \Box f_2  
+ c_4\, (g_{\alpha\beta}\, \partial_\mu\, \partial_\nu\, +
 g_{\mu\nu} \,\partial_\alpha\,\partial_\beta) \Box f_2 \nonumber \\
&+& c_5\,  (g_{\alpha\nu}\, \partial_\mu\, \partial_\beta\, +
g_{\alpha\mu}\, \partial_\nu\, \partial_\beta\, +
g_{\beta\nu}\, \partial_\mu\, \partial_\alpha\, +
g_{\beta\mu}\, \partial_\nu\, \partial_\alpha\ )\, \Box f_2 \biggl] \,.      
\end{eqnarray}
If we insert Eq.~(\ref{eq:f1}) and Eqs.~(\ref{eq:t1})-(\ref{eq:t4})
into this expression and compare with  Eq.~(\ref{eq:2DC1}),
we find five conditions on the five coefficients. The unique solution of 
these conditions gives
\begin{equation}
c_1 = -8\,, \qquad c_2 = -c_4 = -2\,, \quad {\rm and} \quad  c_3 = -c_5 = 1\,.
\end{equation} 
As a check, the  correlation function may be shown explicitly to satisfy
the conservation law 
\begin{equation}
 \partial^\mu\,C_{\mu \nu \alpha \beta}(x,x') =
 \partial^{\alpha'}\,C_{\mu \nu \alpha \beta}(x,x') = 0\,.
\end{equation}

\section{}
\label{sec:AppB}

Here we repeat the derivation in the previous appendix for the case of
four-dimensional Minkowski spacetime. The general form, Eq.~(\ref{eq:C2}),
for the correlation function still holds, but the two-point function
for a massless scalar field is now
\begin{equation}
D = \frac{1}{4 \pi^2\,\Delta x^2} \,. 
\end{equation} 

If we insert this form into Eq.~(\ref{eq:C2}), we find the four-dimensional
analog of Eq.~(\ref{eq:2DC1}):
\begin{eqnarray}
C_{\mu \nu \alpha \beta}(x,x') &=& \frac{1}{4 \pi^4}\, \biggl[
\frac{32}{(\Delta x^2)^6} \; 
 \Delta x_\mu\ \Delta x_\nu \Delta x_\alpha \Delta x_\beta 
       \nonumber \\   
&-& \frac{4}{(\Delta x^2)^5} \; (g_{\mu\alpha}\,\Delta x_\nu\, \Delta x_\beta
+ g_{\mu\beta}\,\Delta x_\nu\, \Delta x_\alpha
+ g_{\nu\alpha}\,\Delta x_\mu\, \Delta x_\beta
+ g_{\nu\beta}\,\Delta x_\mu\, \Delta x_\alpha) 
       \nonumber \\ 
&-& \frac{8}{(\Delta x^2)^5} \; (g_{\mu\nu}\,\Delta x_\alpha\, \Delta x_\beta
 +   g_{\alpha\beta}\,\Delta x_\mu\, \Delta x_\nu)  
    \nonumber \\ 
&+& \frac{1}{(\Delta x^2)^4} \; (g_{\mu\alpha}\, g_{\nu\beta} +
g_{\mu\beta}\, g_{\nu\alpha} +4 g_{\mu\nu}\, g_{\alpha\beta}) \biggr]
                                  \label{eq:4DC1} 
\end{eqnarray}

In four dimensions, the correlation function has dimensions of 
$1/{\rm length}^8$. Thus
any expression involving derivatives on a dimensionless function will
require eight derivatives. Because there are only four free indices,
there will have to be at least two wave operators. This eliminates
the logarithm function $f_1$, Eq.~(\ref{eq:f1def}), because in four dimensions 
\begin{equation}
\Box \Box f_1 = 0\,.
\end{equation} 
However, the squared logarithm function $f_2$ may be used to form the 
following five tensors with the correct dimensions and symmetry:
\begin{equation}
\partial_\mu\, \partial_\nu\,\partial_\alpha\,
                          \partial_\beta\,\Box \Box f_2 \,,
\end{equation} 
 \begin{equation}
(g_{\alpha\beta}\, \partial_\mu\, \partial_\nu\, +
 g_{\mu\nu} \,\partial_\alpha\,\partial_\beta) \Box^3 f_2 \, ,
\end{equation} 
 \begin{equation}
 (g_{\alpha\nu}\, \partial_\mu\, \partial_\beta\, +
g_{\alpha\mu}\, \partial_\nu\, \partial_\beta\, +
g_{\beta\nu}\, \partial_\mu\, \partial_\alpha\, +
g_{\beta\mu}\, \partial_\nu\, \partial_\alpha\ ) \Box^3 f_2 \, ,
\end{equation} 
\begin{equation}
g_{\mu\nu} \,g_{\alpha\beta}\, \Box^4 f_2 \,,
\end{equation}
and
\begin{equation}
(g_{\mu\alpha}\, g_{\nu\beta} + g_{\mu\beta}\, g_{\nu\alpha})\Box^4 f_2 \,.
\end{equation}
We may repeatedly use Eqs.~(\ref{eq:1box}) and (\ref{eq:2box}) with $n=4$ to
demonstrate that, in four-dimensions,
\begin{equation}
\Box \Box f_2 = -\frac{32}{(\Delta x^2)^2} \,,
\end{equation}  
\begin{equation}
\Box^3 f_2 = -\frac{256}{(\Delta x^2)^3} \,,
\end{equation}
and   
\begin{equation}
\Box^4 f_2 = -\frac{6144}{(\Delta x^2)^4} \,.
\end{equation}  
From these expressions, we may show  that
\begin{eqnarray}
\partial_\mu\, \partial_\nu\,\partial_\alpha\,\partial_\beta\,
\Box \Box f_2 &=&
-\frac{61440}{(\Delta x^2)^6} \; 
 \Delta x_\mu\ \Delta x_\nu \Delta x_\alpha \Delta x_\beta 
       \nonumber \\   
&+& \frac{6144}{(\Delta x^2)^5} \; 
(g_{\mu\alpha}\,\Delta x_\nu\, \Delta x_\beta
+ g_{\mu\beta}\,\Delta x_\nu\, \Delta x_\alpha
+ g_{\nu\alpha}\,\Delta x_\mu\, \Delta x_\beta  \nonumber \\
&+& g_{\nu\beta}\,\Delta x_\mu\, \Delta x_\alpha
+ g_{\mu\nu}\,\Delta x_\alpha \Delta x_\beta
+  g_{\alpha\beta}\,\Delta x_\mu\ \Delta x_\nu ) 
       \nonumber \\   
&-& \frac{768}{(\Delta x^2)^4} \; (g_{\mu\alpha}\, g_{\nu\beta} +
g_{\mu\beta}\, g_{\nu\alpha} + g_{\mu\nu}\, g_{\alpha\beta}) \,, 
\end{eqnarray}
\begin{eqnarray}
 & &(g_{\alpha\nu}\, \partial_\mu\, \partial_\beta\, +
g_{\alpha\mu}\, \partial_\nu\, \partial_\beta\, +
g_{\beta\nu}\, \partial_\mu\, \partial_\alpha\, +
g_{\beta\mu}\, \partial_\nu\, \partial_\alpha\ ) \Box^3 f_2 = 
 \frac{3072}{(\Delta x^2)^4} \; (g_{\mu\alpha}\, g_{\nu\beta} +
g_{\mu\beta}\, g_{\nu\alpha}) \nonumber \\
&-& \frac{12288}{(\Delta x^2)^5} \; 
(g_{\mu\alpha}\,\Delta x_\nu\, \Delta x_\beta
+ g_{\mu\beta}\,\Delta x_\nu\, \Delta x_\alpha
+ g_{\nu\alpha}\,\Delta x_\mu\, \Delta x_\beta  
+ g_{\nu\beta}\,\Delta x_\mu\, \Delta x_\alpha) \, ,
\end{eqnarray}
and
\begin{equation}
(g_{\alpha\beta}\, \partial_\mu\, \partial_\nu\, +
 g_{\mu\nu} \,\partial_\alpha\,\partial_\beta) \Box^3 f_2 =
3072 \left[ \frac{ g_{\mu\nu}\, g_{\alpha\beta}}{(\Delta x^2)^4}
-4 \frac{ g_{\mu\nu}\,\Delta x_\alpha \Delta x_\beta
+  g_{\alpha\beta}\,\Delta x_\mu\ \Delta x_\nu}{(\Delta x^2)^5}\right] \, .
\end{equation} 

We now express the correlation function as a sum of the  tensors formed from
$f_2$ as
\begin{eqnarray}
C_{\mu \nu \alpha \beta}(x,x') &=& \frac{1}{61440\, \pi^4}\; \biggl[
c_1\, \partial_\mu\, \partial_\nu\,
                  \partial_\alpha\,\partial_\beta\,\Box \Box f_2
+ c_2\, g_{\mu\nu}\, g_{\alpha\beta}\,  \Box^4 f_2 \nonumber \\
&+& c_3 \, (g_{\mu\alpha}\, g_{\nu\beta} + 
g_{\mu\beta}\, g_{\nu\alpha})\, \Box^4 f_2  
+ c_4\, (g_{\alpha\beta}\, \partial_\mu\, \partial_\nu\, +
 g_{\mu\nu} \,\partial_\alpha\,\partial_\beta) \Box^3 f_2 \nonumber \\
&+& c_5\,  (g_{\alpha\nu}\, \partial_\mu\, \partial_\beta\, +
g_{\alpha\mu}\, \partial_\nu\, \partial_\beta\, +
g_{\beta\nu}\, \partial_\mu\, \partial_\alpha\, +
g_{\beta\mu}\, \partial_\nu\, \partial_\alpha\ )\, \Box^3 f_2 \biggl] \,.
\end{eqnarray}
If we insert the explicit forms for these tensors and compare with
Eq.~(\ref{eq:4DC1}), we again find five conditions on the five coefficients,
leading to the solution
\begin{equation}
c_1 =-8,  \qquad c_2 = -c_4 = -6, \quad {\rm and} \quad  c_3 = -c_5 = -1\,.
\end{equation} 
As required, the correlation function has a vanishing divergence on
any index.

\section{}
\label{sec:AppC}

In this appendix, we will prove Eqs.~(\ref{eq:int_K0}) and (\ref{eq:S_sq})
in both two and four dimensions.
We will proceed by first showing that  
\begin{equation} 
\int_{-\infty}^{\infty} K(t_0) dt_0 = 0 \,. 
\end{equation} 
Then we will prove that $K(t_0)$ is a symmetric function, and  
hence show that  
\begin{equation} 
\int_{0}^{\infty} K(t_0) dt_0 = 0 \,,  
\end{equation} 
as well.  
 
Let $g(t)$ be an arbitrary smooth sampling function.  
From Eq.~(\ref{eq:St_0-S0_a}) or (\ref{eq:4D_worldline}) and 
Eq.~(\ref{eq:Kt_0}), we have  
\begin{equation} 
K(t_0)=\frac{1}{\langle S^2(0) \rangle} \, \int_{-\infty}^{\infty} dt \, 
g(t-t_0) \, \int_{-\infty}^{\infty} dt' \, g(t')\,\, C(t-t') \,. 
\end{equation} 
Then  
\begin{eqnarray}  
\int_{-\infty}^{\infty} K(t_0) dt_0 &=&  
\frac{1}{\langle S^2(0) \rangle} \, \int_{-\infty}^{\infty} dt \, 
\int_{-\infty}^{\infty} dt_0 \, g(t-t_0) \, 
\int_{-\infty}^{\infty} dt' \, g(t')\,\, C(t-t')  \nonumber \\ 
&=& \frac{1}{\langle S^2(0) \rangle} \,\int_{-\infty}^{\infty} dt' \, g(t')\, 
\int_{-\infty}^{\infty} dt \,\,C(t-t') \,, 
\end{eqnarray}  
where we have interchanged the order of integrations, and used the fact that  
for $y=t-t_0$,  
\begin{equation} 
\int_{-\infty}^{\infty} dt_0 \, g(t-t_0)= -\int_{\infty}^{-\infty} dy \, g(y) 
= \int_{-\infty}^{\infty} dy \, g(y) = 1 \,. 
\end{equation} 
However, if we can write $C(t-t') = \partial F(t-t')/\partial t$, where  
$F(t-t') \rightarrow 0$ as $t \rightarrow \pm \infty$,  
then  
\begin{equation} 
\int_{-\infty}^{\infty} dt \,\,C(t-t')= {[F(t-t')]}_{t=-\infty}^{t=+\infty} = 0\,, 
\label{eq:C-condition} 
\end{equation} 
which in turn implies that   
\begin{equation} 
\int_{-\infty}^{\infty} K(t_0) dt_0 = 0 \,. 
\end{equation}  
Recall that in two dimensions, the worldline vacuum correlation function is  
$C(t-t') = 1/[4 \pi^2 {(t-t')}^4]$, and in four dimensions it is
$C(t-t') = 3/[2 \pi^4 {(t-t')}^8]$,  
so in both cases the condition Eq.~(\ref{eq:C-condition}) is satisfied.
Note that in four dimensions, it is necessary to assume that we set the
spatial separation $r$ in Eq.~(\ref{eq:4D_C2}) to zero and 
then average over time,
as discussed in Sect.~\ref{sec:wl_sampling-4D}.  
 
We now show that $K(t_0)=K(-t_0)$. Let us write  
\begin{eqnarray} 
\langle S^2(0) \rangle \, K(t_0) &=& \int_{-\infty}^{\infty}  
dt \, g(t-t_0) \,  \int_{-\infty}^{\infty}  
dt' \, g(t') \, C(t-t')  \nonumber \\ 
&=& \int_{-\infty}^{\infty}  
d{\bar t} \, g({\bar t}) \, \int_{-\infty}^{\infty}  
dt' \, g(t') \, C({\bar t}+t_0-t')  \nonumber \\ 
&=& \int_{-\infty}^{\infty}  
d{\bar t} \, g({\bar t}) \, \int_{-\infty}^{\infty}  
d{\bar t'} \, g({\bar t'}+t_0) \,   
C({\bar t}-{\bar t'}) \,, 
\end{eqnarray} 
where we have let $\bar t = t- t_0$, so $t=\bar t+t_0$,  
and ${\bar t'} = t'- t_0$. If we now let ${\bar t'}  
\rightarrow t, \,\bar t \rightarrow t'$, we have  
\begin{eqnarray} 
\langle S^2(0) \rangle \, K(t_0) &=& \int_{-\infty}^{\infty}  
dt \, g(t+t_0) \, \int_{-\infty}^{\infty}  
dt' \, g(t') \, C(t'-t) \nonumber \\ 
&=& \langle S^2(0) \rangle \, K(-t_0) \,, 
\end{eqnarray} 
where we have used the fact $C(t'-t) = C(t-t')$.  
Note that the symmetry of $K(t_0)$ depends only on  
that of $C$ and does not assume that the sampling function  
$g(t)$ is symmetric. Thus since  
\begin{equation} 
\int_{-\infty}^{\infty} \, K(t_0) \,dt_0=0 \,, 
\end{equation} 
and $K(t_0)$ is symmetric,  
it also follows that  
\begin{equation} 
\int_{0}^{\infty} \, K(t_0) \,dt_0=0 \,. 
\end{equation} 
 
In order to determine whether  
a fluctuation is correlated or anti-correlated with itself,  
we must determine the sign of $\langle S^2(0) \rangle $ in  
the general case. We would expect that a fluctuation should  
be correlated with itself, and thus that $\langle S^2(0) \rangle >0$.  
This can be proven from the fact that $S(0)$, as defined by
Eq.~(\ref{eq:St_0}), is a self-adjoint operator~\cite{Comment}. 
Let $|\psi \rangle$ be the state
under consideration, which in our case is the Minkowski vacuum.
Then
\begin{equation}
|\Psi \rangle = S(0) |\psi \rangle
\end{equation}
is a well defined state vector with positive norm. Thus we have
\begin{equation}
||\Psi ||^2 = \langle \psi | S^\dagger(0)\, S(0) |\psi \rangle
=\langle S^2(0) \rangle > 0 \,.
\end{equation}


\begin{thebibliography}{28}
\bibitem{Comment} A similar argument was used after Eq.~(2.10) in 
Ref.~\cite{MV99}. We would like to thank Albert Roura for drawing our 
attention both to the argument and to this reference.
\end{thebibliography}
\end{document}